\documentclass[aps,prd,epsf,showpacs,amsmath,amssymb,graphics,onecolumn,preprint,12pt]{revtex4}
\usepackage{graphicx}
\usepackage{dcolumn}
\usepackage{bm}

\begin{document}
\title{\bf Particle acceleration by Majumdar-Papapetrou di-hole}

\author{$^1$Mandar Patil\footnote{Electronic address: mandar@iucaa.ernet.in},
$^2$Pankaj S. Joshi\footnote{Electronic address: psj@tifr.res.in},
}

\affiliation{$^1$Inter University Center for Astronomy and Astrophysics \\ Post Bag 4, Ganeshkhind, Pune-411007, India. \\
$^2$Tata Institute of Fundamental Research, Homi Bhabha Road, Mumbai 400005, India.
}

\begin{abstract}
We explore the multi-black hole spacetimes from the perspective of the ultra-high energy particle collisions. 
Such a discussion is limited to the spacetimes containing a single black hole so far.
We deal with the Majumdar-Papapetrou solution representing a system consisting of two identical black holes 
in the equilibrium. In order to identify the conditions suitable for the process of high energy collisions,
we consider particles confined to move on the equatorial plane towards the axis of symmetry with the zero angular 
momentum. We consider collision between the particles moving in opposite directions at the location midway between 
the black holes on the axis. We show that the center of mass energy of collision between the particles increases 
with the decrease in the separation between the black holes and shows divergence in the limit where the separation 
goes to zero. We estimate the size of the region close to the central point on the equatorial plane where it would 
be possible to have high energy collisions and show that this region has a reasonably large spatial extent. 
We further explore the process of high energy collisions with the general geodesics with arbitrary angular 
momentum on the 
equatorial plane away from the central point. Although in this paper we deal with the Majumdar-Papapetrou 
spacetime which serves as a toy example representing multiple black holes, we speculate on the possibility that 
the ultra-high energy collisions would also occur in the more general setting like colliding black holes, when
distance between the black holes is extremely small, which can in principle be verified in the numerical
relativity simulations.

\end{abstract}


\maketitle

\section{Introduction}

Terrestrial particle accelerators like Large Hadron Collider can accelerate and collide particles with the
center of mass energy of collision upto 10 TeV. The physics beyond this energy scale all the way up to the 
Planck scale remains 
unexplored experimentally. An intriguing possibility is to make use of the naturally occurring 
extreme gravity scenarios like black holes to facilitate the high energy collisions and extract information 
about the physics at
that scale by observing various particles produced in the collision events. Basic idea is to make use of gravity 
instead
of electromagnetic forces to 'accelerate' the particles.

Such process was first proposed in the context of extremally rotating
Kerr black hole by Banados, Silk and West \cite{BSW}. They considered two particles dropped in from rest at 
infinity moving
along the timelike geodesics on the equatorial plane of the Kerr black hole. They considered collision between 
the particles
in the vicinity of event horizon of the black hole. It was shown that the center of mass
energy of collision shows divergence in the limit where the black hole is close to being
extremal i.e. Kerr spin parameter is arbitrarily close to unity and angular momentum of one of the particles
takes a specific finetuned value. Various aspects of this process were analyzed by different authors \cite{BH1}. 
A similar process 
in the context of numerous other examples of black holes including Reissner-Nordstr\"{o}m black hole \cite{Zasla} 
was also studied \cite{BH2}. 
But the analysis so far is confined to the spacetimes that contain a single black hole. 
In this paper we show that the high energy collisions can also take place in 
the spacetimes that contain two black holes instead of one and this collision process is qualitatively different from
the conventional Banados-Silk-West(BSW) process.

Various drawbacks of the BSW process were also pointed out \cite{Berti},\cite{Jacobson},\cite{Piran},\cite{Mc}. 
The Kerr spin parameter has to be finetuned to a value that is arbitrarily close to unity to achieve large center of 
mass energy of collision \cite{Jacobson}. The finetuning 
of the angular momentum of one of the particles is required \cite{BSW,Berti,Jacobson}. The proper time required 
for the collision to occur is also
infinite in the reference frame attached to the particle with finetuned angular momentum \cite{BSW,Berti,Jacobson}.
The fraction of the collision
products that escape to infinity is vanishingly small \cite{Mc}. The super-heavy exotic particles produced in 
the collisions would not escape to infinity \cite{Piran}.
Conservative backreaction of the colliding particles could put a severe limitation on the highest 
center of mass energy of collision achievable in the collision \cite{Nakao,Zhu}.

We had explored the process of high energy collisions in the context of naked singularities \cite{Patil1,Patil2,Patil3}.
It was qualitatively different from the BSW process around the black holes.
It was shown that the collision between an ingoing particle
and another ingoing particle turned into an outgoing particle either due to the repulsive effect of gravity or 
due to its angular momentum, could be arbitrarily large. 
But again it was necessary to finetune the spin parameter to the extremal value. Various drawbacks 
associated with the BSW process in the context of the black holes were however avoided. 
Finetuning of the angular
momentum of colliding particles was not required \cite{Patil1}. The proper time required for the collision 
to occur was finite \cite{Patil1,Patil3}. The escape fraction of the collision products was finite \cite{Patil4}. 
The center of mass energy of collision in a reasonably realistic 
astrophysical scenario after taking into account the conservative backreaction was shown to be finite but larger 
than the Planck
energy \cite{Patil3}. We further showed that 
neither black hole nor naked singularity was necessary for the high energy collisions to occur \cite{Patil5,Patil6}.
But again the solutions 
where high energy collisions could be realized were near extremal in some suitable sense.

In this paper we explore the binary black hole solutions from the perspective of the high energy collisions.
The two body problem in general relativity is extremely complicated unlike that in Newtonian gravity 
essentially due to the complexity of the Einstein equations. Thus there is no realistic solution available 
in the literature representing the binary black holes. So we work in the context of a particular simplistic 
solution to the Einstein equations representing the equal mass binary black holes.
This solution was obtained by Majumdar and Papapetrou \cite{Majumdar,Papapetrou} and was later interpreted to 
be the spacetime representing multiple black holes \cite{Hawking}. The solution is static. In addition to that there
is a rotational symmetry about the line joining two black holes and a
reflection symmetry about the plane passing through the point midway between the black holes and which is perpendicular
to the axis. All these symmetries simplify analysis of the geodesic motion of the particle in the spacetime while 
calculating the center of mass energy of collision. Initially in order to identify the conditions suitable 
for the ultra-high energy collisions to occur we consider a simple scenario where particles are restricted 
to move along the 
equatorial plane and move towards the axis. We show that the center 
of mass energy of collision between the particles that move in the opposite direction and collide at 
the point midway between
the black holes goes on increasing with the decrease in the separation between the black holes  
and shows divergence in the limit where the separation goes to zero. We also estimate the size of 
the region which would
host the high energy collisions and show that it could be reasonably large. We then deal with the general scenario 
where particles with the non-zero angular momentum participate in the high energy collisions away from the 
central point 
on the equatorial plane.

In the realistic scenarios representing binary black hole system these symmetries would be
absent. So it will be interesting to analyze whether the process of high energy collisions would still occur 
in the region between the two black holes when they are sufficiently close to one another for instance in a scenario 
like binary black hole coalescence. This analysis is impossible to carry out analytically due to the absence of 
exact solution. 
However this can in principle be implemented in the numerical relativity simulation of the   colliding black holes.

\section{Majumdar-Papapetrou solution}

In this section we describe the Majumdar-Papapetrou solution representing multiple black holes.
We first carry out analysis in the isotropic rectangular coordinates and deal with the simplistic 
scenario in order to identify the conditions that must be satisfied so that it would be possible to 
have ultra-high energy collisions. Later we use the isotropic cylindrical coordinates which are 
suitable for the more general analysis.

In Newtonian mechanics a set of particles each with the charge same as the mass can be 
in the equilibrium for arbitrary separation between the charges. This is because the 
electrostatic repulsion between the particles is balanced by the gravitational attraction.
Similar situation also arises in the context of general relativity. A set of extremally 
charged black holes each with the charge same as the mass and located at arbitrary locations 
that are in the equilibrium appear as the solution to the Einstein-Maxwell equations and are 
described by the Majumdar-Papapetrou solution \cite{Majumdar,Papapetrou}.

Majumdar-Papapetrou solution in the isotropic rectangular coordinates is given by 
\begin{equation}
 ds^2=-\frac{1}{U^2}dt^2+U^2\left(dx^2+dy^2+dz^2\right)
\end{equation}
where 
\begin{equation}
 U=1+\sum_{i}\frac{M_i}{\sqrt{(x-x_i)^2+(y-y_i)^2+(z-z_i)^2}}
\end{equation} 
Black holes of masses $M_i$ are situated at the locations $(x_i,y_i,z_i)$. In this coordinate system each black hole 
is represented by just a point. However one can carry out an analytical extension and show that 
$(x_i,y_i,z_i)$ represents an event horizon with area $4\pi M_{i}^{2}$.
The gauge field is given by 
\begin{equation}
 A_{\mu}=\left(U^{-1},0,0,0\right)
\end{equation}

We are interested in the situation where Majumdar-Papapetrou solution represents two identical black holes
separated by some distance. We assume that both the black holes are located along $z-$axis at the coordinates 
given by $(x_1=0,y_1=0,z_1=a)$ and
$(x_2=0,y_2=0,z_2=-a)$. Thus there is a rotation symmetry about $z-$axis and also there is a reflection 
symmetry about $z=0$ plane.
The metric is given by 
\begin{equation}
 ds^2=-\frac{1}{U^2}dt^2+U^2\left(dx^2+dy^2+dz^2\right)
\end{equation}
with
\begin{equation}
 U=1+\frac{M}{\sqrt{x^2+y^2+(z-a)^2}}+\frac{M}{\sqrt{x^2+y^2+(z+a)^2}}
\end{equation} 
where $M$ is a mass of each of the black holes.

In order to identify the conditions to be imposed on the spacetime geometry necessary to have
ultra-high energy collisions we deal with the simple scenario where the colliding particles move 
in the equatorial plane $z=0$ towards the axis of symmetry. Such particles have zero angular momentum.
In order to deal with this situation we use Majumdar-Papapetrou metric written in the isotropic 
rectangular coordinate system.

Later on in order to analyze the collisions between particles with non-zero angular momenta we
use the isotropic cylindrical coordinate system which is well-suited for this purpose. We make the 
following coordinate transformation
\begin{equation}
 x=\rho \cos{\phi}~,~ y=\rho \cos\phi
\end{equation}
Majumdar-Papapetrou metric in the isotropic cylindrical coordinate system can be written as 
\begin{equation}
 ds^2=-\frac{1}{U^2}dt^2+U^2\left(d\rho^2+\rho^2 d\phi^2+dz^2\right)
\end{equation}
where
\begin{equation}
 U=1+ \frac{M}{\sqrt{\rho^2+(z-a)^2}}+\frac{M}{\sqrt{\rho^2+(z+a)^2}}
\end{equation}

There are many interesting investigations carried out recently in the context of the Majumdar-Papapetrou 
equal mass di-hole spacetime. The shadow of the binary di-hole system was calculated and was shown to be 
significantly different from the mere superposition of shadows of the individual black holes \cite{Shadow}.
As the black hole are brought closer there is a deformation of shadow of individual black holes. Shadows 
turn elliptical from being spherical. In addition to that there appears an eyebrow like structure outside the 
main shadows. Width of the eyebrow depends on the distance between the black holes. The appearance of the eyebrow 
is due to the light rays which can go around and suffer a large deflection from both the black holes while
traveling from source to the observer.

The structure of the circular timelike and null geodesics in the equatorial plane was investigated in \cite{circular}. 
It was shown that the existence of null circular orbit crucially depends on the separation between the black holes 
for a fixed mass. If the black holes are far apart then there are no circular null orbits. If the black holes 
are close enough
there are two circular null orbits. Whereas for the specific critical separation between the black holes there 
is only one degenerate circular orbit. Timelike circular orbit can exist everywhere except for the region
between the photon sphere. The circular orbits for unequal mass binary black holes also exhibit interesting features.

In this paper we explore a different aspect of the Majumdar-Papapetrou spacetime. 
We investigate the process of high energy collisions between the particles for specific 
trajectories whose analysis is simplified due to the existence of symmetries and try get 
insights regarding what might be expected in the more general and realistic scenarios.

\section {Geodesics motion in the isotropic rectangular coordinates}
In this section we describe motion of the particles that follow geodesics in the Majumdar-Papapetrou 
spacetime in the rectangular isotropic coordinates.
We assume that the particles are massive although a similar discussion will follow
for the massless particles as well. The effects of the conservative and dissipative backreaction on 
the particle motion are ignored. In order to understand the conditions that must be imposed on 
the spacetime geometry that would lead to the high energy collisions we deal with a simple 
scenario that we describe below.

Let $U^{\mu}=\left(U^t,U^x,U^y,U^z\right)$ be four-velocity of the particle.
There is a reflection symmetry about $z=0$ plane. As a consequence of which we expect that 
the particle which is moving
on this plane will not leave this plane and continue to travel on it. This can be shown in the following way.
The geodesic equation for the motion of the particle along $z$ direction is given by

\begin{eqnarray}
\nonumber
&&\frac{d U^z}{\text{d$\tau $}} + \frac{\left(\frac{M (-a+z)}{\left(x^2+y^2+(a-z)^2\right)^{3/2}}+\frac{M (a+z)}{\left(x^2+y^2+(a+z)^2\right)^{3/2}}\right)}{\left(1+\frac{M}{\sqrt{x^2+y^2+(a-z)^2}}+\frac{M}{\sqrt{x^2+y^2+(a+z)^2}}\right)^5}U^{\text{t2}} 
+\frac{\left(\frac{M (-a+z)}{\left(x^2+y^2+(a-z)^2\right)^{3/2}}+\frac{M (a+z)}{\left(x^2+y^2+(a+z)^2\right)^{3/2}}\right)} {\left(1+\frac{M}{\sqrt{x^2+y^2+(a-z)^2}}+\frac{M}{\sqrt{x^2+y^2+(a+z)^2}}\right)}U^{\text{x2}}\\ 
&&+\frac{\left(\frac{M (-a+z)}{\left(x^2+y^2+(a-z)^2\right)^{3/2}}+\frac{M (a+z)}{\left(x^2+y^2+(a+z)^2\right)^{3/2}}\right)} {\left(1+\frac{M}{\sqrt{x^2+y^2+(a-z)^2}}+\frac{M}{\sqrt{x^2+y^2+(a+z)^2}}\right)}U^{\text{y2}} 
-\frac{\left(\frac{M (-a+z)}{\left(x^2+y^2+(a-z)^2\right)^{3/2}}+\frac{M (a+z)}{\left(x^2+y^2+(a+z)^2\right)^{3/2}}\right)} {\left(1+\frac{M}{\sqrt{x^2+y^2+(a-z)^2}}+\frac{M}{\sqrt{x^2+y^2+(a+z)^2}}\right)}U^{\text{z2}}\\\nonumber
&&-\frac{2\left(\frac{M x}{\left(x^2+y^2+(a-z)^2\right)^{3/2}}+\frac{M x}{\left(x^2+y^2+(a+z)^2\right)^{3/2}}\right)}{\left(1+\frac{M}{\sqrt{x^2+y^2+(a-z)^2}}+\frac{M}{\sqrt{x^2+y^2+(a+z)^2}}\right)}U^xU^z 
-\frac{2\left(\frac{M y}{\left(x^2+y^2+(a-z)^2\right)^{3/2}}+\frac{M y}{\left(x^2+y^2+(a+z)^2\right)^{3/2}}\right)} {\left(1+\frac{M}{\sqrt{x^2+y^2+(a-z)^2}}+\frac{M}{\sqrt{x^2+y^2+(a+z)^2}}\right)}U^yU^z=0
\end{eqnarray}
\newline

If the particle is initially in the $z=0$ plane and if it is moving along $z=0$ plane i.e. if $U^z=0$, 
then it can be seen using the equation
above that $\frac{d U^z}{\text{d$\tau $}}=0$. Thus the particle will not acquire velocity along $z$ axis 
and will continue to travel in the equatorial plane $z=0$.

There is a rotational symmetry about the axis of symmetry i.e. $z$ axis. Thus if 
the particle is moving in the equatorial plane along the line passing through the axis of symmetry, it will continue 
to move along that line as we show it below. For simplicity we consider a particle
that is moving along $x$ axis. 

The geodesic equation for the motion of the particle along $y$ axis is given by

\begin{eqnarray} \nonumber
&&\frac{\text{dU}^y}{\text{d$\tau $}} + \frac{\frac{M y}{\left(x^2+y^2+(a-z)^2\right)^{3/2}}+\frac{M y}{\left(x^2+y^2+(a+z)^2\right)^{3/2}}}{\left(1+\frac{M}{\sqrt{x^2+y^2+(a-z)^2}}+\frac{M}{\sqrt{x^2+y^2+(a+z)^2}}\right)^5}U^{\text{t2}}
+\frac{\left(\frac{M y}{\left(x^2+y^2+(a-z)^2\right)^{3/2}}+\frac{M y}{\left(x^2+y^2+(a+z)^2\right)^{3/2}}\right) }{\left(1+\frac{M}{\sqrt{x^2+y^2+(a-z)^2}}+\frac{M}{\sqrt{x^2+y^2+(a+z)^2}}\right)}U^{\text{x2}}\\ 
&&-\frac{\left(\frac{M y}{\left(x^2+y^2+(a-z)^2\right)^{3/2}}+\frac{M y}{\left(x^2+y^2+(a+z)^2\right)^{3/2}}\right) }{\left(1+\frac{M}{\sqrt{x^2+y^2+(a-z)^2}}+\frac{M}{\sqrt{x^2+y^2+(a+z)^2}}\right)}U^{\text{y2}}
+\frac{\left(\frac{M y}{\left(x^2+y^2+(a-z)^2\right)^{3/2}}+\frac{M y}{\left(x^2+y^2+(a+z)^2\right)^{3/2}}\right) }{\left(1+\frac{M}{\sqrt{x^2+y^2+(a-z)^2}}+\frac{M}{\sqrt{x^2+y^2+(a+z)^2}}\right)}U^{\text{z2}}\\\nonumber
&&-\frac{2\left(\frac{M x}{\left(x^2+y^2+(a-z)^2\right)^{3/2}}+\frac{M x}{\left(x^2+y^2+(a+z)^2\right)^{3/2}}\right) }{\left(1+\frac{M}{\sqrt{x^2+y^2+(a-z)^2}}+\frac{M}{\sqrt{x^2+y^2+(a+z)^2}}\right)}U^yU^x
-\frac{2\left(\frac{M (-a+z)}{\left(x^2+y^2+(a-z)^2\right)^{3/2}}+\frac{M (a+z)}{\left(x^2+y^2+(a+z)^2\right)^{3/2}}\right) }{\left(1+\frac{M}{\sqrt{x^2+y^2+(a-z)^2}}+\frac{M}{\sqrt{x^2+y^2+(a+z)^2}}\right)}U^yU^z=0
\end{eqnarray}

We would like to restrict to the case where the particle is moving in the equatorial plane. 
Thus setting $z=0$ and $U^z=0$ we get

\begin{eqnarray}
\nonumber
\frac{\text{dU}^y}{\text{d$\tau $}}&+&\frac{\frac{2M y}{\left(x^2+y^2+a^2\right)^{3/2}}}{\left(1+\frac{2M}{\sqrt{x^2+y^2+a^2}}\right)^5}U^{\text{t2}}+\frac{\frac{2M y}{\left(x^2+y^2+a^2\right)^{3/2}}}{\left(1+\frac{2M}{\sqrt{x^2+y^2+a^2}}\right)}U^{\text{x2}}  \\ 
&-&\frac{\frac{2M y}{\left(x^2+y^2+a^2\right)^{3/2}}}{\left(1+\frac{2M}{\sqrt{x^2+y^2+a^2}}\right)}U^{\text{y2}}+\frac{\left(\frac{4M x}{\left(x^2+y^2+a^2\right)^{3/2}}\right) }{\left(1+\frac{2M}{\sqrt{x^2+y^2+a^2}}\right)}U^yU^x=0 
\end{eqnarray}

Now if the particle is initially located on $x$ axis and is moving along $x$ axis i.e. if $y=0$ and $U^{y}=0$ then it
can be seen from the expression above that $\frac{d U^y}{\text{d$\tau $}}=0$. Thus the particle will not
acquire any velocity 
along $y$ direction and will continue to move along $x$ axis.

Remaining components of the four-velocity can be determined by invoking the static nature of Majumdar-Papapetrou 
spacetime.  
It admits a Killing vector $\partial_t$. Associated with this Killing vector there is a constant of 
motion $ E=-\partial_t \cdot U $ which has an interpretation
of conserved energy of the particle per unit mass. Using this one can write down $U^t$ as
\begin{equation} 
U^t=E\left(1+\frac{2M}{\sqrt{x^2+a^2}}\right)^2
\end{equation}
Using the equation above and the normalization condition for velocity $U \cdot U=-1$ we write down the $U^x$ as
\begin{equation} 
U^x=\pm \sqrt{E^2-\frac{1}{\left(1+\frac{2M}{\sqrt{x^2+a^2}}\right)^2}}
\end{equation}
This equation can be recast in the form 
\begin{equation}
 U^{x2}+V_{eff}(x)=E^2
\end{equation}
where $V_{eff}(x)$ can be interpreted as the effective potential for motion in $x$ direction and is 
given by 
\begin{equation}
 V_{eff}(x)=\frac{1}{\left(1+\frac{2M}{\sqrt{x^2+a^2}}\right)^2}
\end{equation}
It admits a minimum at origin and maximum at infinity
\begin{equation}
 V_{min}=V_{eff}(x=0)=\frac{1}{\left(1+\frac{2M}{a}\right)^2} ~~~V_{max}=V_{eff}(x\rightarrow \infty)=1
\end{equation}
Thus we must have 
\begin{equation}
 E\ge \frac{1}{\left(1+\frac{2M}{a}\right)}
\end{equation}
If $\frac{1}{\left(1+\frac{2M}{a}\right)}<E<1$ then the particle is bound and oscillates back and forth between 
\begin{equation}
 x=\pm \sqrt{\left(\frac{2M}{\frac{1}{E}-1}\right)^2-a^2}
\end{equation}
If $E \ge 1$, particle does not admit turning points and motion of the particle is unbound. Particle with 
$E=1$ starts from rest at infinity.

This is a discussion of the geodesic motion of the particles relevant for the description of process of 
high energy particle
collision around the Majumdar-Papapetrou binary black holes. As we shall describe in the next section we 
consider two particles that move along the $x$ axis in the opposite direction and undergo collision at 
the origin.We compute the center of mass energy of collision and analyze its variation with the 
separation between the black holes. 

\section{Ultra-high energy collisions in isotropic rectangular coordinates}

In this section we describe the process of ultra-high energy particle collisions around 
di-hole in a simple scenario in order to understand the conditions that must be imposed 
on the spacetime. As described earlier a particle initially moving on the equatorial plane 
along the line directed towards 
the axis of symmetry will continue to move along this line as a consequence of rotational 
symmetry around the line joining 
two black holes and reflection symmetry about the equatorial plane. 

We consider two identical particles each with the mass $m$ moving along $x$ axis in the 
opposite directions. One along positive $x$ direction and 
other along negative $x$ direction. For simplicity we assume that they have same conserved energy $E$.  
We consider the collision between these particles at the arbitrary value of $x$ in the allowed region. 

Non-zero components of the four-velocities of two particles $U_{1}^{\mu}$ and $U_2^{\mu}$ are given by 
\begin{eqnarray}
\nonumber
U_{1}^{t}=E\left(1+\frac{2M}{\sqrt{x^2+a^2}}\right)^2, U_{1}^{x}= \sqrt{E^2-\frac{1}{\left(1+\frac{2M}{\sqrt{x^2+a^2}}\right)^2}} \\ \nonumber
U_{2}^{t}=E\left(1+\frac{2M}{\sqrt{x^2+a^2}}\right)^2, U_{2}^{x}= -\sqrt{E^2-\frac{1}{\left(1+\frac{2M}{\sqrt{x^2+a^2}}\right)^2}}
 \end{eqnarray}
The center of mass energy of collision between the particles is given by the expression \cite{BSW}
\begin{equation}
 E_{c.m.}^2=2m^2\left(1-U_1 \cdot U_2\right)
\end{equation}
It is a natural generalization of notion of the center of mass energy defined in special 
relativity that we use while dealing with particle physics and quantum field theory on flat spacetime.
In this case it turns out to be 
\begin{equation}
 E_{c.m.}^2(x)=4m^2E^2 \left(1+\frac{2M}{\sqrt{x^2+a^2}}\right)^2
\end{equation}
The center of mass energy goes on increasing with the decreasing value of $x$ and 
is maximum at the center $x=0$ and its value is given by 
\begin{equation}
 E_{c.m.}^2(x=0)=4m^2E^2 \left(1+\frac{2M}{a}\right)^2
\end{equation}
Clearly the center of mass energy at the origin increases 
with the decreasing separation between the black holes $a$.
It shows divergence in the limit where the distance between the black holes goes to zero.
\begin{equation}
 \lim_{a \to 0}  E_{c.m.}^2(x=0) \sim \frac{1}{a^2} \rightarrow \infty
\end{equation}

So when the black holes are placed close enough, the center of mass energy of collision
in the region between them is extremely large.  

In this section we identified the condition that must be imposed on the spacetime geometry in 
order to have collisions with the large center of mass energy of collisions. We considered the 
collisions that take place between the particles that move on the equatorial plane with zero angular
momentum. Later we shall deal with the more general scenario where the collisions take place between
the particles with the non-zero angular momentum in the region away from the center.

\section{Estimation of size of region hosting high energy collisions}
In this section we estimate size of the region hosting high energy collisions for a typical situation.
We assume that the conserved energy of the particles is $E=1$ and mass of the colliding particles is 
approximately equal to the mass of proton $m \simeq 1 GeV$. 

The center of mass energy of collision is maximum at the center 
\begin{equation}
 E_{e.m.}(x=0)\simeq \frac{4mM}{a}
\end{equation}
In the region where $a<<x<<M$ the center of mass energy of collision is given by 
\begin{equation}
 E_{e.m.}(x)\simeq \frac{4mM}{x}
\end{equation}

Suppose the center of mass energy of collision between the two particles at the center 
is $10^4$ times larger than the LHC scale which is $1TeV=10^4GeV$ i.e.
\begin{equation}
E_{c.m.}\simeq 10^8 GeV
\end{equation}
which occurs when the separation between the black holes is given by 
\begin{equation}
 \frac{M}{a}\simeq 4 \times 10^8 
\end{equation}

The location away from the center upto which it would be possible to have collisions
with the energies achievable at LHC
\begin{equation}
 E_{c.m.}(x_{LHC})\simeq 10^4 GeV
\end{equation}
is given by 
\begin{equation}
 x\simeq 10^4a= 10^{-4}M
\end{equation}

For a supermassive black hole of mass $M\simeq10^8M_{sun}$ 
the size of the region which can host collisions with energies larger than the LHC scale is given by 
\begin{equation}
 x_{LHC}\simeq 10^4 M_{sun} \simeq 3 \times 10^4 km 
\end{equation}

Thus size of the region around center which hosts collisions with high center of mass energies is can be 
reasonably large.

\section{Geodesic motion in isotropic cylindrical coordinate system }
In the earlier sections we dealt with the geodesic motion in the equatorial plane with the zero 
angular momentum. In this section we deal with the particles that move in the equatorial plane with
the non-zero angular momentum. We use the isotropic cylindrical coordinates for this purpose. 
The metric can be written as 

\begin{equation}
 ds^2=-\frac{1}{U^2}dt^2+U^2\left(d\rho^2+\rho^2 d\phi^2+dz^2\right)
\end{equation}
where
\begin{equation}
 U=1+ \frac{M}{\sqrt{\rho^2+(z-a)^2}}+\frac{M}{\sqrt{\rho^2+(z+a)^2}}
\end{equation}

The geodesic equation for the motion of the particle along $z-$axis is given by
\begin{eqnarray}
\nonumber
\frac{\text{dU}^z}{\text{d$\tau $}} &+& \frac{\left(\frac{M (z-a)}{\left((z-a)^2+\rho ^2\right)^{3/2}}+\frac{M (a+z)}{\left((z+a)^2+\rho
^2\right)^{3/2}}\right)}{\left(1+\frac{M}{\sqrt{(z-a)^2+\rho ^2}}+\frac{M}{\sqrt{(z+a)^2+\rho ^2}}\right)^5}U^{\text{t2}}+\frac{\left(\frac{M (z-a)}{\left((z-a)^2+\rho
^2\right)^{3/2}}+\frac{M (a+z)}{\left((z+a)^2+\rho ^2\right)^{3/2}}\right)}{\left(1+\frac{M}{\sqrt{(z-a)^2+\rho ^2}}+\frac{M}{\sqrt{(z+a)^2+\rho
^2}}\right)}U^{\text{$\rho $2}} \\ 
\nonumber
&+&\frac{\rho ^2\left(\frac{M (z-a)}{\left((z-a)^2+\rho ^2\right)^{3/2}}+\frac{M (a+z)}{\left((z+a)^2+\rho ^2\right)^{3/2}}\right)}{\left(1+\frac{M}{\sqrt{(z-a)^2+\rho
^2}}+\frac{M}{\sqrt{(z+a)^2+\rho ^2}}\right)}U^{\text{$\phi $2}}-\frac{\left(\frac{M (z-a)}{\left((z-a)^2+\rho ^2\right)^{3/2}}+\frac{M (a+z)}{\left((z+a)^2+\rho
^2\right)^{3/2}}\right)}{\left(1+\frac{M}{\sqrt{(z-a)^2+\rho ^2}}+\frac{M}{\sqrt{(z+a)^2+\rho ^2}}\right)^5}U^{\text{z2}}\\
&+&\frac{2M \rho \left(\frac{M (z-a)}{\left((z-a)^2+\rho ^2\right)^{3/2}}+\frac{M (a+z)}{\left((z+a)^2+\rho ^2\right)^{3/2}}\right)}{\left(1+\frac{M}{\sqrt{(z-a)^2+\rho
^2}}+\frac{M}{\sqrt{(z+a)^2+\rho ^2}}\right)}U^{\phi }U^z=0
\end{eqnarray}
It is clear from the expression above that since $z=0$ and $U^z=0$, it turns out that $\frac{dU^z}{dz}=0$. 
Thus the particle which is initially moving in $z=0$ plane continues to stay on this plane. 

For simplicity we deal with the particles that are constrained to move on the equatorial 
plane $z=0$ as it would be impossible to deal with the off-equatorial motion in absence 
of the Carter-like constant since the Hamilton-Jacobi equation is not integrable. 

It is quite clear from the form of the metric that it admits Killing vectors
$\partial_{t}$ and $\partial_{\phi}$ that correspond to the static nature of spacetime and cylindrical symmetry
about the line joining two black holes. Associated with these Killing vectors there are 
constants of motion $E=-\partial_{t} \cdot U$ and $L= \partial_{\phi} \cdot U$ which can be 
interpreted as conserved energy and angular momentum per unit mass of the particle. Using these 
one can write down the following components of four-velocity of the particle in the isotropic 
cylindrical coordinate system. 

\begin{equation}
U^{t}=\Omega^2 E~~,~~U^{\phi}=\frac{L}{\Omega^2\rho^2}
\end{equation}
where
\begin{equation}
 \Omega=\left(1+\frac{2M}{\sqrt{\rho^2+a^2}}\right)
\end{equation}
Using the equations above and the normalization condition for the four-velocity $U \cdot U=-1$
we can write down the radial component of velocity as 
\begin{equation}
 U^{\rho}=\pm \sqrt{E^2-\frac{1}{\Omega^2}\left(1+\frac{L^2}{\Omega^2\rho^2}\right)}
\end{equation}
This equation can be rewritten as 
\begin{equation}
 U^{\rho 2}+V_{eff}=E^2
\end{equation}
where 
\begin{equation}
 V_{eff}= \frac{1}{\Omega^2}\left(1+\frac{L^2}{\Omega^2\rho^2}\right)
\end{equation}
can be thought of as an effective potential for the radial motion.

From now onwards we express all the quantities in units of $M$ i.e. we set $M=1$.
We also assume that $a<<1$ since we have demonstrated earlier that it would be possible to have
collisions with ultra-high center of mass energies if separation between the black holes is extremely small.
We now intend to explore collisions between the particles in the more general setting when the particles 
have non-zero angular momenta 

Let 
\begin{equation} 
\rho= \alpha a
\end{equation}

We are interested in the region close to the center where we expect high energy collisions 
would take place. 

Let 
\begin{equation}
 \alpha << \frac{1}{a}
\end{equation}
So that 
\begin{equation}
\Omega^2 \simeq \frac{4}{(1+\alpha^2)a^2}
\end{equation}

The approximate value of the effective potential turns out to be 
\begin{equation}
 V_{eff}\simeq \frac{(a+\alpha^2)a^2}{4}\left(1+\frac{L^2 (1+\alpha^2)}{4\alpha^2}\right)
\end{equation}

For the finite angular momentum and $\alpha \sim O(1)$ it turns out that, $V_{eff}\simeq a^2 <<E^2$ and 
therefore the particle has a finite radial velocity component. 

Thus it is necessary to have $\alpha <<1$ for initially ingoing particle to turn back as an 
outgoing particle. In that case it turns out that 
\begin{equation}
V_{eff}\simeq \frac{a^2}{4}\left(1+\frac{L^2}{4\alpha^2}\right)=E^2
\end{equation}
We get 
\begin{equation}
 \alpha \simeq \frac{1}{\sqrt{\frac{4}{L^2}\left(\frac{4E^2}{a^2}-1\right)}}\simeq \frac{aL}{4E}
\end{equation}
Thus the ingoing particle with energy $E$ and angular momentum $L$ will turn back into an outgoing 
particle at the location radial coordinate $\rho=\alpha a \simeq \frac{a^2L}{4E}$.
Collisions between the particles with non-zero angular momentum would take place at the location 
away from the center as these particles cannot reach the center.

The derivative of the effective potential at the turning point is given by 
\begin{equation}
 \frac{dV_{eff}}{dr}\simeq -\frac{4E^2}{a}
\end{equation}
which takes a nonzero value.

In the next section we describe the process of collisions between the particles with the non-zero angular momenta. 

\section{Ultra-high energy particle collisions in isotropic cylindrical coordinate system}
In this section we deal the collisions between the particles moving in the equatorial plane with the 
non-zero angular momenta away from the center in the cylindrical coordinate system.

We consider two particles with the identical values of conserved energy $E_1=E_2=E$ and absolute value 
of the angular momenta $|L_1|=|L_2|=L$.

Four-velocities of the two particles are given by 
\begin{eqnarray}
\nonumber
 U^{t}_1=\Omega^2 E~~,~~U^{\phi}_1=\nu_{1}\frac{L}{\Omega^2\rho^2}~~,
~~ U^{\rho}_1=\beta_1 \sqrt{E^2-\frac{1}{\Omega^2}\left(1+\frac{L^2}{\Omega^2\rho^2}\right)} \\
 U^{t}_2=\Omega^2 E~~,~~U^{\phi}_2=\nu_{2}\frac{L}{\Omega^2\rho^2}~~,
~~ U^{\rho}_2=\beta_2 \sqrt{E^2-\frac{1}{\Omega^2}\left(1+\frac{L^2}{\Omega^2\rho^2}\right)} 
\end{eqnarray}
where $\nu_1,\nu_2=\pm 1$ depending on whether the angular momentum is oriented parallel or 
anti-parallel to the $z-$axis and $\beta_1,\beta_2=\pm 1$ depending on whether the particle 
is moving radially outwards or inwards. 

We assume that 
\begin{equation}
 \frac{aL}{4E} \le \alpha << \frac{1}{a}
\end{equation}
i.e. the collision takes place at the radial location 
\begin{equation}
 \frac{a^2L}{4E} \le \rho << 1
\end{equation}

The center of mass energy of collision between the particles is given by 
\begin{eqnarray}
\nonumber
E_{c.m.}^2 &&=2m^2\left(1-U_{1}\cdot U_{2}\right) \\
&&=2m^2\left( \left(1+\beta_1\beta_2\right)+\left(1-\beta_1\beta_2\right)E^2\Omega^2+\left(\beta_1\beta_2 - \nu_1\nu_2\right)\frac{L^2}{\rho^2\Omega^2}\right)
\end{eqnarray}

\subsection{{\bf Case 1. $\beta_1\beta_1=1~,~\nu_1\nu_2=1$}}
We deal the case where both the particles are either radially ingoing or 
outgoing and the angular momenta of both the particles are oriented either parallel or 
anti-parallel to $z-$axis. Thus they have identical four-velocities and they are at rest with 
respect to each other. It can be easily verified that the center of mass energy will be simply 
the addition of their rest mass energies. 
\begin{equation}
 E_{c.m.}^2=4m^2
\end{equation}
 
\subsection{{\bf Case 2. $\beta_1\beta_1=-1~,~\nu_1\nu_2=-1$}}
In this case one of the particles is radially ingoing and the other particle is radially outgoing.
The angular momenta of one of the particles is parallel and that of the other particle is antiparallel to the $z-$axis.
\begin{equation}
 E_{c.m.}^2=4m^2E^2\Omega^2\simeq \frac{16E^2}{a^2(1+\alpha^2)}
\end{equation}
Close to the turning point where 
\begin{equation}
\alpha <<1
\end{equation}
the center of mass energy is given by 
\begin{equation}
 E_{c.m.}^2=\frac{16E^2}{a^2}
 \end{equation}
which is extremely large when the separation between the black holes is extremely small.
\begin{equation}
 E_{c.m.}^2 \rightarrow \infty
\end{equation}

Whereas when 
\begin{equation}
 1<< \alpha << \frac{1}{a}
\end{equation}
the center of mass energy goes as 
\begin{equation}
 E_{c.m.}^2\simeq \frac{16E^2}{\rho^2} 
\end{equation}
This situation is similar to the one encountered in section V. Thus the center of mass energy will be 
large in the region of reasonable size around the center.

\subsection{{\bf Case 3. $\beta_1\beta_1=1~,~\nu_1\nu_2=-1$}}
In this case both the particles either move in the radially inward or radially outward direction. 
Whereas angular momentum of one of the particles is oriented parallel and that of the other 
particle is oriented anti-parallel to the $z-$axis.
\begin{equation}
 E_{c.m.}^2=2m^2\left(2+\frac{2L^2}{\rho^2\Omega^2}\right) \simeq 2m^2\left(2+\frac{L^2(1+\alpha^2)}{2\alpha^2}\right) 
\end{equation}
Let 
\begin{equation}
 \alpha=\sigma \left(\frac{L a}{4E}\right)~~,~~\sigma \ge 1
\end{equation}
We get 
\begin{equation}
 E_{c.m.}^2 \simeq 2m^2\left(2+\frac{8E^2}{a^2\sigma^2}\left(1+\frac{\sigma^2 a^2 L^2}{16E^2}\right)\right) 
\end{equation}
As long as 
\begin{equation}
 \sigma << \frac{1}{a}
\end{equation}
we get
\begin{equation}
 E_{c.m.}^2 \simeq \frac{8E^2}{a^2\sigma^2} 
\end{equation}
Thus the center of mass energy of collision is large if the distance between the black holes
is extremely small.
\begin{equation}
 E_{c.m.}^2 \rightarrow \infty
\end{equation}
In this case the center of mass energy is large only if the collision takes place close to the turning point.

\subsection{{\bf Case 4. $\beta_1\beta_1=-1~,~\nu_1\nu_2=+1$}}
Here one of the particles moves radially inwards and the other particle moves outwards
and angular momenta of both the particles are oriented either parallel or anti-parallel 
to the $z-$axis.

The center of mass energy of collision is given by 
\begin{equation} 
E_{c.m.}^2=4m^2\left(E^2\Omega^2-\frac{L^2}{\rho^2\Omega^2}\right)\simeq  \frac{16m^2E^2}{a^2(1+\alpha^2)}-\frac{m^2L^2(1+\alpha^2)}{\alpha^2}
\end{equation}
Let 
\begin{equation}
 \alpha= \sigma \frac{1}{\sqrt{\frac{4}{L^2}\left(\frac{4E^2}{a^2}-1\right)}}~~,~~\sigma\ge 1
\end{equation}
For 
\begin{equation}
 \sigma \simeq O(1)
\end{equation}
we get 
\begin{equation}
 E_{c.m.}^2\simeq \frac{4m^2}{\sigma^2}+ \frac{16 m^2E^2}{a^2}\left(1-\frac{1}{\sigma^2}\right)
\end{equation}
Very close to the turning point both the particles have identical four-velocities and thus the center
of mass energy of collision is twice the rest mass.
\begin{equation}
 \sigma \rightarrow 1~~,~~E_{c.m.}^2 \simeq 4m^2 
\end{equation}
whereas sufficiently away from the turning point we get 
\begin{equation}
 E_{c.m.}^2\simeq \frac{16 m^2E^2}{a^2}\left(1-\frac{1}{\sigma^2}\right)
\end{equation}
It is clear from the expression above that the center of mass energy of collision is extremely large.
\begin{equation}
 E_{c.m.}^2 \rightarrow \infty
\end{equation}

Whereas when 
\begin{equation}
 1<< \alpha << \frac{1}{a}
\end{equation}
we get 
\begin{equation}
 E_{c.m.}^2\simeq \frac{16E^2}{\rho^2} 
\end{equation}
which is again similar to the situation encountered in section V.
Thus the center of mass energy of collision will be large in the finite region of reasonable size. 

It is clear from the analysis of the four cases above that whenever one of the colliding particle is 
radially ingoing and other one is radially outgoing it is possible to have collisions with the large 
center of mass energy of collisions in the region of reasonably large size as long as two black holes 
are close to one another. 

Whereas when both the particles are either radially ingoing or outgoing the center of mass 
energy of collision will be large in the region close to the turning point if their angular 
momenta are oriented antiparallel.

\section{Comparison with single black hole}

The process of the high energy collision in the context of the binary black holes is quite different 
compared to the analogous process in the context of a single black hole \cite{BSW,Zasla}. 
The major qualitative difference
is as given below.

In case of the extremally spinning black hole, the collision is between the two ingoing particles in the vicinity
of the event horizon. One of the particles must have the critical angular momentum. This particle asymptotically 
approaches the event horizon as the particle reaches the maximum of the effective potential for the radial motion.
Its velocity approaches zero as it goes near the horizon.
As a consequence of which the proper time required for this particle to reach the event horizon 
and participate in the 
collision is infinite \cite{BSW,Jacobson}.

Similarly in the context of the extremally charged 
Reissner-Nordstrom black hole the collision is between particle with a critical charge 
and another neutral particle in the 
vicinity of the event horizon.
A particle with the critical charge asymptotically approaches the horizon since it 
approaches the maximum of the effective
potential. Its radial velocity approaches zero and the proper time required is again infinite 
as in the case of the extremal Kerr
black hole \cite{Zasla,Nakao}. 

As far as the binary black holes are concerned the situation is quite different.
In the first case that we dealt with the situation where the colliding particles have zero angular momenta 
and both the particles have finite velocity in the $x$ direction
at the collision point $x=0$. This point is not a maximum but rather a minimum of the effective potential.
Thus the proper time required for the particles to reach the desired location and participate in the
collision is finite unlike in the single black hole case. Particles with the non-zero angular momenta 
turn back at the location close to the center where the effective potential equals the square of 
conserved energy but the derivative of the effective potential is non-zero thus again the proper 
time required is finite.

Thus the process of high energy collision in the context of the binary black holes
is qualitatively different from the Banados-Silk-West process in the context of the single black holes.

\section{Will collision process generalize to realistic binary black holes ?}

We described the process of high energy collisions in the background of the Majumdar-Papapetrou equal mass binary 
black holes. It is an exact solution to the electro-vacuum Einstein equations representing the binary black holes.
It admits certain spacetime symmetries. Using symmetries of the spacetime it becomes possible to analyze the
geodesic motion 
and identify the geodesics relevant for the process of high energy collision.

Majumdar-Papapetrou solution however is not very realistic. Since we expect that in the realistic scenario will 
have the binary black holes that would be spinning and uncharged which would be moving around in the space. Thus 
all symmetries of the Majumdar-Papapetrou metric would be absent in the realistic scenario. 
Majumdar-Papapetrou spacetime however serves 
as a toy model in the context of which calculations of the relevant process can be done and intuition for what might 
happen in the general setting can be derived.

Based on the calculation in the Majumdar-Papapetrou spacetime we arrive at the conclusion that it
would be possible to have
collisions with extremely large center of mass energies in the region between the black holes when they are close 
to each other. Similar results might continue to hold good in case of the other more realistic binary black holes.
Since the two body problem in general relativity is extremely difficult to deal with, because of the complexity 
of the Einstein equations, we do not have any exact solutions representing the realistic binary black hole systems.
However numerical simulations of the evolution of the binary black holes are available. 
It will be interesting to see whether 
such collisions can be realized in the numerical relativity simulations when the black holes are close to each other 
on the verge of merger.

Supermassive black holes at the center of the galaxies are expected to interact and form binary 
system during the event of the merger
of their host galaxies (\cite{BBH1} and references therein). They might come closer due to the their 
interaction with the nearby stars and gas. 
When they are sufficiently close, they are expected to evolve predominantly due to the emission of the gravitational
waves. They loose energy and 
angular momentum due to the emission of the gravitational waves and come closer during the inspiral phase and 
eventually collide and 
merge. We expect to detect the gravitational waves emitted from the binary black hole system. If the results 
obtained in this paper
also generalize to the realistic binary black hole systems then the colliding black holes could provide us 
with the naturally
occurring scenarios where the high energy collisions could take place.

\section{Conclusions}
In this paper we describe the process of the ultra-high energy collisions in the background of the binary black holes.
Such a discussion was so far confined to the case of single black holes.
We deal with the subcase of the Majumdar-Papapetrou multi black hole 
electro-vacuum solution representing 
two identical black holes separated by some distance.
We consider particles that are confined to move along the equatorial
plane towards the midpoint between the black holes along the axis of symmetry.
The collision is between the particles
moving in the opposite directions at the point midway between the black holes. We compute the center of 
mass energy of collision.
We show that the center of mass energy 
of collision goes on increasing as the separation between the black holes is decreased and shows divergence when 
the separation goes to zero. We also generalize this process to the particles with the non-zero angular momentum 
that can participate in the high energy collisions away from the central point on the equatorial plane. 
The size of the region which 
can host the ultra-high energy collisions is shown to be reasonably large. Thus it is possible to have 
ultra-high energy collisions in the region between the two
black holes when they are close to each other. It will be interesting to see whether this result would 
generalize to the
more realistic examples of the binary black hole systems such as those that appear in the numerical relativity 
simulations
of the colliding black holes. If these results generalize, then the colliding black holes would provide us 
with the naturally
occurring scenarios where the ultra-high energy collisions could be realized.

\end{document}